# 3D Printed Lightweight Composite Foams

Bharath H S[1], Dileep Bonthu[1], Pavana Prabhakar[2] and Mrityunjay Doddamani[1*]

[1]Advanced Manufacturing Laboratory, Mechanical Engineering, National Institute of Technology, Karnataka, Surathkal, India.

[2]Department of Civil and Environmental Engineering, University of Wisconsin-Madison, Madison, WI, 53706, USA

*mrdoddamani@nitk.edu.in

*All authors contributed equally.*

## Abstract

The goal of this paper is to enable 3D printed lightweight composite foams by blending hollow glass micro balloons (GMB) with high density polyethylene (HDPE). To that end, lightweight feedstock for printing syntactic foam composites is developed. The blend for this is prepared by varying GMB content (20, 40, and 60 volume %) in HDPE for filament extrusion, which is subsequently used for three-dimensional printing (3DP). The rheological properties and the melt flow index (MFI) of blends are investigated for identifying suitable printing parameters. It is observed that the storage and loss modulus, as well as complex viscosity, increases with increasing GMB content, whereas MFI decreases. Further, the coefficient of thermal expansion of HDPE and foam filaments decreases with increasing GMB content, thereby lowering the thermal stresses in prints, which promotes the reduction in warpage. The mechanical properties of filaments are determined by subjecting them to tensile tests, whereas 3D printed samples are tested under tensile and flexure tests. The tensile modulus of the filament increases with increasing GMB content (8-47%) as compared to HDPE and exhibit comparable filament strength. 3D printed foams show higher specific tensile and flexural modulus as compared to neat HDPE, making them suitable candidate materials for weight sensitive applications. HDPE having 60% by volume GMB exhibited the highest modulus and is 48.02% higher than the printed HDPE. Finally, the property map reveals higher modulus and comparable strength against injection and compression molded foams. Printed foam registered 1.8 times higher modulus than molded samples. Hence, 3D printed foams have the potential for replacing components processed through conventional manufacturing processes that have limitations on geometrically complex designs, lead time, and associated costs.

## Nomenclature

| | | | |
|---|---|---|---|
| $\rho$ | Density | $T_{Cryst}$ | Crystallization temperature |
| $\rho_c$ | Composite density | $T_{Melt}$ | Peak melting temperature |
| $\rho_f$ | Density of Filler | $\eta'$ | Complex viscosity |
| $\rho_m$ | Density of Matrix | $G'$ | Storage modulus |
| $\rho_{th}$ | Theoretical density | $G''$ | Loss modulus |
| $\rho_{exp}$ | Experimental density | $\phi_f$ | Cenospheres volume % |
| $V_f$ | Filler volume % | $\phi_v$ | Void content |
| $V_m$ | Matrix volume % | $\alpha_{Cryst}$ | Degree of crystallinity |

## Introduction

Traditional manufacturing of thermoplastic based closed cell foams is realized through injection or compression molding processes [1, 2]. These methods require tooling for fabricating complex parts, which can be expensive and time consuming. However, research on additive manufacturing (AM) indicates that FFF (fused filament fabrication) is among the widely utilized technique to create complex functional parts [3]. Further, AM eliminates the standard constraints on the component size along with producing highly complex parts with zero tooling cost, lower energy, and material consumption [4-6]. Although most polymers are currently used in FFF based 3DP, the development of lightweight thermoplastic filaments for specialized applications is still in its infancy. Thermoplastic composites are used in semi-structural and many engineering applications as they are environmentally friendly and offer flexibility of processing using various methods [7]. Commonly used thermoplastic polymers like polymethylmethacrylate [8], polylactide [9], acrylonitrile butadiene styrene [10, 11], polycarbonate [12], polyetherimide [13] filaments produced from their respective blends [14, 15] are used in industrial 3D printers as feedstock material. Polymers like high-density polyethylene (HDPE) [16], polypropylene [17], polyamide [18], polycaprolactone [19], polybutylene terephthalate [20] etc. have limited studies due to warpage and delamination associated issues that can be addressed by adding various inorganic/organic fillers using compounding methods.

Inorganic and organic solid fillers have been used extensively in thermoplastic industries [21-24]. Reinforcing filler particles in the matrix has several benefits, including a reduction in resin costs as well as flexibility in tailoring properties [25]. Mechanical, surface, electrical, and magnetic properties can be altered using such fillers [2, 26]. The most commonly used fillers are $Al_2O_3$ [27], glass [28], iron particles [29], carbon and glass fibers [30]. Hollow spherical particles like fly ash and GMBs as fillers in the matrix are investigated earlier using conventional processing methods wherein higher tooling cost and complex geometrical design restrictions pose many challenges [31-34]. Closed-cell composite foams (hollow microballoons reinforced in the matrix) provide greater versatility in designing underwater vehicle structures, including internal descriptions of instrumentation housings, buoyancy chambers [35], etc. By changing the volume % of these hollow fillers in the matrix, tailor-made properties can be achieved for many different applications [26, 36]. Achieving these properties depends on particle survival in these lightweight foams and processing methods used for synthesizing them [2, 37]. Developing lightweight filament with minimum to no particle breakage should significantly enhance specific properties in 3DP of components for weight sensitive applications like in nose cones of remotely operated underwater vehicles or even printing the entire body in the tubular design form with all the internal structural details all at once. Automotive and aerospace components without any joints, if realized through printing, can add structural stability with enhanced performance. Adhesive joints are the weakest entities in the structure as the pressurization/depressurization leads to foams fracture in marine environment. 3DP of foams can eliminate adhesive bonding of multiple blocks making them to work well in deep sea environments. In order to manufacture complex shapes and contours and eliminate the need for adhesive bonding, foams printing and associated development of specialized lightweight filaments is the need of the time for marine, automotive and aerospace components.

The feedstock filament development poses processability challenges due to density differences between the constituents, filler dispersion, and rheological behavior [38]. Further, developed composite filaments must be in the desired diameter to be fed into commercially available 3D printers with sufficient flexibility for spooling [39]. These properties allow the filaments to be printed through the printer nozzle without any breakage leading to a block-free layered deposition of prints with dimensional stability [28]. Similarly, the prints quality rendered by the FFF is governed by various parameters like extrusion temperature, the temperature of bed and nozzle, the orientation of print, percentage of infill (filling of the space), layer height, and raster width [40].

In addition, semi-crystalline polymer printing includes issues like shrinkage/warpage, build plate adhesion, and post-print removal [28, 41]. Adding thermally stable inorganic fillers to semi-crystalline polymers minimize shrinkage and make products dimensionally stable [42]. The composite components, on the other hand, show a considerable variation in thermal properties and experience distinct thermal cycles during subsequent processing. Hence, careful choice of processing temperatures and cooling rates ensures quality prints [43], and realizing such lightweight foams is crucial. Reinforcing matrix with hollow fillers results in the reduction of the matrix volume % leading to lightweight composite structures known as syntactic foams. These closed cell composite foams may be categorized into two, three, and multi-phase systems depending on the different types of phases present. They have better mechanical properties and can produce complex functional parts that can replace HDPE, thereby lowering carbon footprints [44-46]. Naturally available fly ash cenospheres have numerous surface defects [47] as compared to engineered microballoons, and hence engineered GMBs are chosen in the current work. Introducing hollow GMB particulate fillers in a matrix can impart significant weight reduction and can be effectively exploited for weight sensitive structures. Weight (density) and strength are essential material properties and are crucial for aeronautical, naval, and automotive components. In manufacturing low cost lightweight thermoplastics without compromising mechanical properties of the material, GMBs are candidate filler exhibiting promising behavior [2]. GMB based foams shown to have better mechanical properties than fly ash based composite foams [7, 48]. Owing to better mechanical responses and biocompatibility, HDPE finds its application in chemical containers, milk jugs, household utilitarian, and other structural applications [49, 50]. Replacing HDPE with engineered GMBs and realizing 3D printed lightweight complex structure having enhanced specific mechanical properties is of paramount interest for weight sensitive structural applications. Nonetheless, for such applications, hollow particle survival, warpage, layer adhesion needs to be carefully investigated and is the focus of this paper.

GMB based 3D printed foam structures can be effectively used in many structural applications [51-53] owing to higher specific mechanical properties, as mentioned earlier. The inclusion of such stiffer GMB particles in HDPE matrix can make the prints more dimensionally stable [54]. The load carrying capacity of 3D printed parts depends on the infill % [3, 55]. In this paper, GMB/HDPE blends are prepared with varying filler content and are tested for melt flow index and rheology. Filaments extruded from the blend are investigated under DSC and tensile tests. Extruded lightweight filaments are then fed into a 3D printer, and the fabricated prints are investigated under DSC, CTE, rheology, flexural, and tensile tests. Finally, a property map is plotted to compare the 3D printed GMB/HDPE composite foam results with other composites realized using conventional fabrication processes. Such a comparison act as a guiding tool for material selection based on specific end-product requirements.

**Experimental**
*Materials*
Hollow GMB's (iM30k, 3M corporation, Singapore) are used as fillers, with an average diameter of 15.3 μm, density of 0.6 gm/cm$^3$, crushing strength of 27000 psi, and a wall thickness of 1.4 μm. HDPE (HD50MA180) used as the matrix is purchased from IOCL, Mumbai, India (Table 1) with a 3 mm mean granule size. GMBs are varied at 20, 40, and 60% by volume in H (HDPE matrix) and are designated as H20, H40, and H60, respectively. GMB content lower than 20% results in non-uniform dispersion in the HDPE matrix, while higher than 60% results in increased viscosity of the melt, leading to microballoon breakage as observed in the pilot experiments. Figure 1 shows the micrographs of GMB and HDPE used in the current work. Smooth surface without any surface defects is observed for both GMB (Figure 1a) and HDPE (Figure 1b). Further, GMB particles are spherical in shape, aiding uniform resin spread during processing.

Table 1. Typical Characteristics of HDPE granules*.

| Property | Typical Value |
|---|---|
| Melt flow index | 20 gm/10 min |
| Density | 0.950 gm/cm$^3$ |
| Flexural Modulus | 750 MPa |
| Vicat Softening Point | 124°C |
| Tensile Strength @ Yield | 22 MPa |
| Elongation @ Yield | 12 % |

*As mentioned by the supplier.

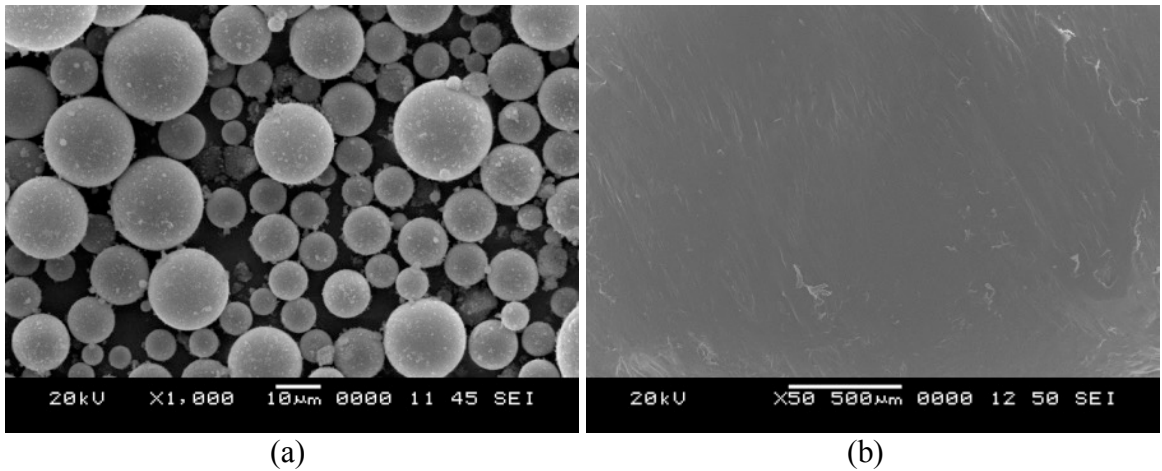

(a) (b)
Figure 1. Micrographs of as received (a) GMB and (b) HDPE.

*Preparation of Blend, MFI, and rheological properties*
A Brabender (16CME SPL) is used for blending HDPE and GMB. Blending speed and temperature are set at 10 rpm and 160°C, respectively [2, 26] to prepare H20, H40, and H60 compositions. The representative image of the H60 blend is shown in Figure 2a. Dynisco LMI5000 melt flow index equipment is used for measuring MFI (ASTM D1238) of H - H60 pellets, which helps in setting an appropriate multiplier in printing by isolating different temperature settings for different compositions. The study of rheological properties is essential to know the effect of filler on manufacturing conditions. Anton Paar rotational rheometer, MCR 502, is used to investigate the influence of fillers on the rheology of the developed blends. A 25 mm diameter and 1mm thick specimen are used for frequency sweep of 0.1 to 10 Hz at 150 °C at a 5% loading rate.

*Filament development and 3D printing*
The extrusion process is carried using a 25SS/MF/26 single screw extruder supplied by Aasabi machinery Pvt. Ltd., Bombay with an L/D ratio (flight length of screw to its outside diameter) of 25:1. The composite blends are pre-heated at 80 °C for 24 hours to eliminate the moisture, if any, before gravity feeding into the extruder hopper. Foam pellets (Figure 2a) are fed into an extruder having a barrel temperature profile of 145-150-155-145 °C (feed - die segment). The screw speed is set at 25 rpm. Take-off unit speed is maintained at 11.5 rpm to extrude the filaments of 2.85±0.05 mm in diameter (Figure 2b).

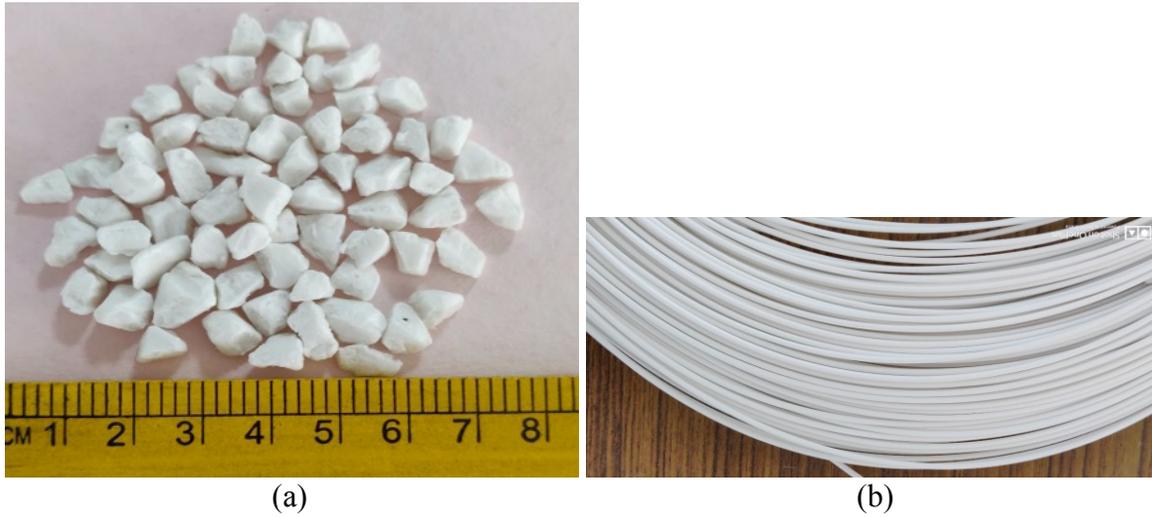

(a) (b)
Figure 2. Representative (a) blend of GMB/HDPE and (b) extruded H60 feedstock filament.

Obtained H - H60 filaments are used as input material for 3D printing. Commercially available FFF based Star 3D printer supplied by AHA 3D Innovations Pvt. Ltd., Jaipur has two nozzles of 0.5 mm diameter. The suitable values of temperature and flow rates based on the pilot experiments are set to achieve completely rigid parts with 100% infill for comparing with fully dense molded components. Printing at higher temperatures can help to achieve temperature distribution uniformly alongside the annealing effect yielding in a better adhesion of layers and dimensionally stability. Nozzle and bed temperatures below 225 and 80 °C respectively resulted in improper material flow through nozzles and non-uniform bonding of the raft with the HDPE plate, which is placed on the glass bed of the printer. Higher material flow through the nozzle and HDPE plate distortion are observed for the temperatures above 245 and 100 °C, respectively, for nozzle and printer chamber. The experimental strategy followed for identifying suitable printing parameters based on layer deposition, defects, layer adhesion, post printing removal and warpage are discussed in the later section. Samples are printed on the HDPE plate. After printing, samples are left on the build plate until it reaches room temperature to minimize the warpage. Later, prints are used for characterization.

*DSC and CTE investigations*
Perkin Elmer DSC-6000, USA, is used to estimate melting and crystallization on filaments and prints of H - H60 compositions. 10 mg of the specimen is heated in a 30 μl Al crucible for 0-200 °C temperature range with isothermal curing at 200 °C for about 3 min. Later, samples are brought to zero degrees at a rate of 10 degrees/minute, eliminating thermal history due to earlier processing steps. Post cooling at 0 °C for 3 min, the samples are heated again from 0-200 °C. DSC plots display endothermic and exothermic peaks representing melting enthalpy at cold crystallization. Crystallinity % ($\alpha_{Cryst}$) is assessed as,

$$\alpha_{Cryst} = \frac{\Delta H_m}{\Delta H_m^*} \times 100 \qquad (1)$$

where, $\Delta H_m$= heat of fusion in J/g and $\Delta H_m^*$= heat of fusion/gram for HDPE, 293 J/g [56]. Dilatometer, CIPET, Chennai, is used to estimate CTE for prints (ASTM D696-13) having a dimension of 75×12.7×3 mm. CTE values qualitatively exhibit warpage and dimensional stability information [57].

*Void content and density estimations*
According to ASTM D792-13, filaments and prints experimental densities are calculated. Using the rule of mixture, the theoretical density is determined by,
$$\rho_c = V_f \rho_f + V_m \rho_m \qquad (2)$$
where, *m, f, c, V,* and *ρ* are the matrix, filler, composite, volume fraction, and density, respectively. The difference in theoretical and experimental densities gives % void content and is given by [58],
$$\emptyset_v = \frac{\rho_{th} - \rho_{exp}}{\rho_{th}} \qquad (3)$$
Such matrix porosity (void) in prints implies raster gaps though the infill is 100%. These air gaps developed while printing leads to three-phase foam structures helping in enhancing the energy absorbing capabilities.

*Tensile and flexural investigations*
Filament and 3D printed samples are tensile tested using Zwick Roell Z020, USA, with a 20 kN load cell. The total length of the filament is 176 mm, with a 76 mm distance between the grips. The test is carried out by maintaining a constant 5 mm/min loading rate. An extensometer (gauge length 50 mm) is used to measure the strain. Printed samples are tested according to ASTM D638-14, at similar crosshead displacement using a 25 mm extensometer gauge length. The initial load elongation of 0.1 MPa is recorded using the extensometer. For flexural testing of prints (127 × 12.7 × 3.2 mm), a three-point bending configuration (ASTM D790-17) is adopted with a preload of 0.1 MPa, loading rate of 1.37 mm/min with span length to depth ratio of 16:1. Flexural modulus is computed using,
$$E_f = \frac{L^3 m}{4bd^3} \qquad (4)$$
where, *d: thickness*, *b: width*, *m: slope*, *L: span length*.

Flexural stress ($\sigma_{fm}$) is calculated using,
$$\sigma_{fm} = \frac{3PL}{2bd^2} \quad (P: \text{Load}) \qquad (5)$$
A minimum of five samples are tested for all the experimental investigations, and the average values are reported. Micrographs of as fabricated freeze fractured and post test filaments and prints are taken by gold sputter covering (JFC-1600) using JSM 6380LA JEOL, Japan.

## Results and Discussion
*MFI and rheology of GMB/HDPE*
Flowability is quantified by MFI. An increase in GMB content reduces MFI due to filler resistance to the flow of polymer [59]. HDPE has recorded the highest MFI (17.94 gm/10 min) when compared with H20 (13.76), H40 (8.11), and H60 (4.85). MFI decreased by 23.29, 54.79, and 72.97 %, with increasing GMB by 20, 40, and 60 volume %, respectively. A similar observation is also noted in Ref. [59, 60]. Decreased MFI needs to be carefully looked into either by raising the temperature of printing or increasing the print extrusion multiplier, especially for foams with higher filler loadings. The printing temperature is kept constant for H - H60 to consolidate the warpage, and hence multiplier factor is changed for higher GMB %. An increase in filler infusion increases melt viscosity of the polymer [61] and is observed in the entire frequency sweep (Figure 3a). At higher frequency, HDPE shows a shear-thinning region. H20 - H60 shows similar behavior with a slight increase in $\eta'$ and is due to restriction of polymer chain movements by GMBs. Among foams, H60 shows the highest $\eta'$. At 0.1 and 50 rad/sec, complex viscosities for H, H20, H40 and H60 are in the range of 1080.52 - 636.75, 2045.4 - 1048, 2729.6 - 1324.2 and 4331.4 - 1701.5 Pa-s respectively. Compared to H (11808 Pa at 50 rad/sec), foams have higher storage modulus owing to presence of greater number of stiffer particles (Figure 3b). Storage modulus increases from 20,019 - 32,163 Pa for H20 - H60 foams. HDPE and H20 display standard homopolymer-like terminal behavior at lower frequencies due to the complete relaxation of polymer chains [62]. Compared to pure HDPE, H20 has a higher modulus. Plateau region is observed at a lower frequency for H40 and H60, indicating viscoelasticity. G″ increases with increasing frequency and filler content for all the samples (Figure 3c). The loss modulus for H - H60 ranges between 107.56 - 429.56 Pa, respectively, at 0.1 rad/sec, which is ~4 times for H60 as compared to H. Such a multifold increase in G″ could be due to restrained matrix flow around stiff intact GMBs. Rheological and MFI properties act as a guideline for selecting appropriate process parameters for printing of quality components.

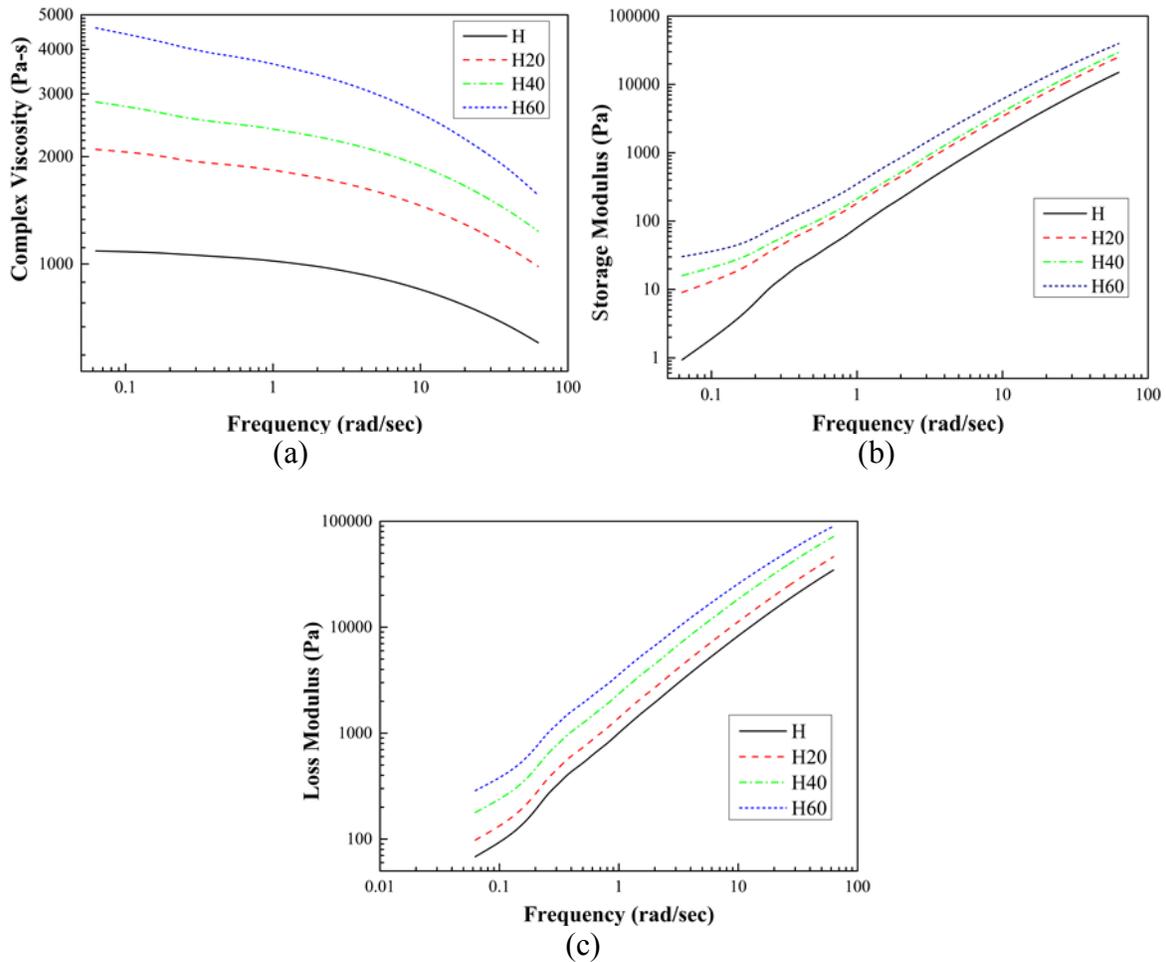

Figure 3. Rheological properties of H - H60 blends.

*Physical and microstructural characterization*

Performance and behavior of extruded foam filaments are influenced by the interaction of filler-matrix, filler %, and matrix porosity. For filaments to be used in a 3D printer, adequate spooling stiffness and strength are needed. Hence, tests to find the density, morphology of extruded filament, and tensile properties are performed before printing to check the quality, stiffness, and strength necessary for filament feasibility to be used in a commercially available printer. Table 2 presents density estimations, void %, and the weight reduction potential of filaments and prints.

Table 2. Physical properties of filament (F) and prints (Pnt).

| Composition | $\phi_f$ (vol. %) | $\rho_{th}$ (kg/m³) | $\rho_{exp}$ (kg/m³) | | $\phi_v$ (%) | | Weight saving potential (%) | |
|---|---|---|---|---|---|---|---|---|
| | | | F | Pnt | F | Pnt | F | Pnt |
| H | 0 | 950 | 942±8 | 927±12 | 0.84 | 2.42 | --- | --- |
| H20 | 20 | 880 | 858±15 | 826±13 | 2.50 | 6.14 | 8.92 | 10.90 |
| H40 | 40 | 810 | 780±11 | 746±18 | 3.70 | 7.90 | 17.20 | 19.53 |
| H60 | 60 | 740 | 683±12 | 668±10 | 7.70 | 9.73 | 27.49 | 27.94 |

The experimental and theoretical densities of HDPE filaments are very close, indicating lower void formations due to its hydrophobic nature. Mechanical properties of HDPE and foams are influenced by voids presence, as an effective load bearing area reduces. An increase in GMB content increases void content in filaments (0.84-7.70%) and prints (2.42-9.73%). Higher void content in print as compared to filaments indicate matrix porosity is transferred from the filament to prints.

Further, additional porosity of 1.58, 3.64, 4.2, and 2.03% is observed in H, H20, H40, and H60 prints, respectively. Such additional porosity in prints amid 100% infill is due to air gaps between the raster. These additional porosities form three phase (HDPE, GMB, and raster gap) syntactic foams enhancing damping capabilities further. As the developed GMB/HDPE filaments are quite flexible, they did not break for freeze fractured micrography amid 24 hours of dipping them in liquid nitrogen. Hence, all the foam filaments are knife cut, and micrographs are shown in Figure 4 to check the cross-section and dispersion of GMBs in the HDPE matrix.

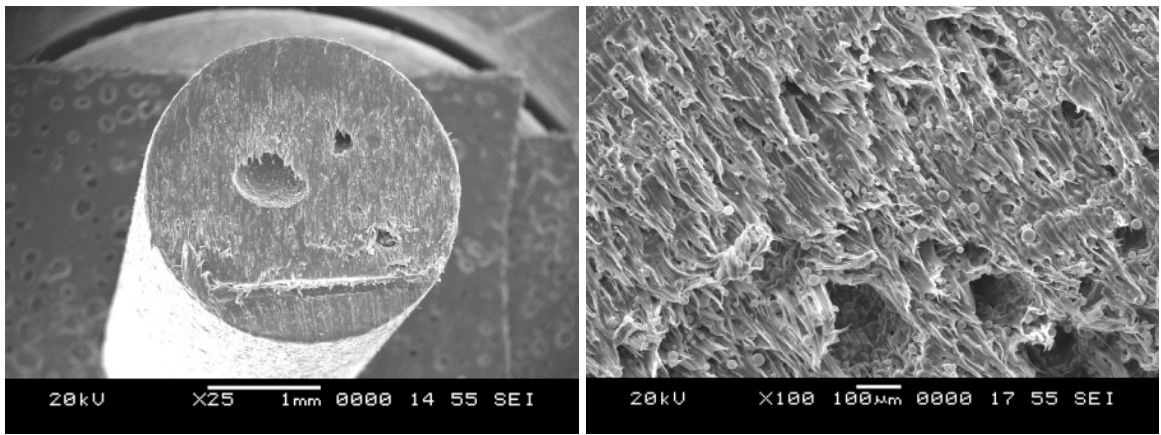

(a)　　　　　　　　　　　　　　　　　(b)

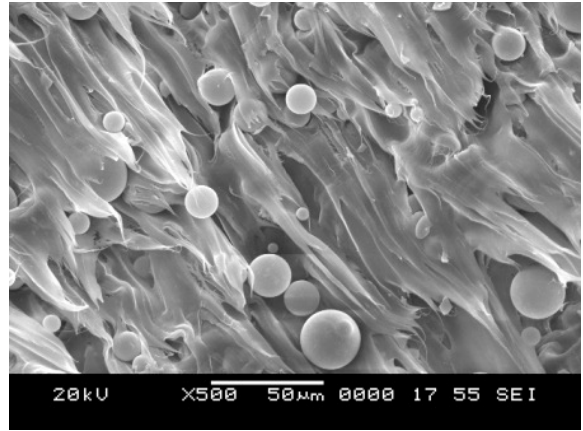
(c)

Figure 4. Extruded filament micrograph of (a) cross sectional view for representative H20. H60 at (b) lower and (c) higher magnification.

The circular cross-section in Figure 4a of representative H20 filament affirms the suitability of chosen extrusion parameters. Figure 4b shows a low magnification micrograph of H60, showing the uniform distribution of intact GMB particles and few voids in compliant HDPE matrix. Such pores/voids, if transferred during 3D printing, may increase three phase syntactic foams compliance resulting in higher damping. Poor interfacial bonding between GMB and HDPE is clearly evident from a higher magnification micrograph of H60 (Figure 4c) and is obvious as constituent materials are used without any surface treatment to avoid additional processing time, cost and difficulty in correlating properties with inconsistently coated layer thickness.

*DSC investigations of filaments and prints*
Thermal behavior ($T_{Cryst}$, $T_{Melt}$, and CTE) of H - H60 is presented in Table 3 for looking into warpage related issues. DSC plots for H - H60 are presented in Figure 5. For pure HDPE, the endothermic peak is observed at 108 °C, which is noted to be in an increasing trend for foams. The decrease in the level of endotherm and crystallization temperature rise with higher GMB content is also noted in Figure 5. This strongly affirms the fact that, while HDPE cools, the nucleation of melt occurs on the filler surface at relatively higher temperatures forming thicker crystal lamellas leading to higher $T_{Crst}$ [63]. Melt inertia is ignored as the crystallization temperature of foams varies in a very narrow range of 2.2% as compared to H (Table 3). An increase in filler volume % has an insignificant influence on $T_{Melt}$ of both filaments and prints, as seen from Table 3 indicating i) additional thermal history imposed by 3DP post extrusion has not induced higher residual thermal stresses and ii) printing temperature can be kept constant for all the samples. A decrease in $\alpha_{Cryst}$ (56.68%) for foam filaments is observed with increasing GMB content relative to HDPE.

Table 3. Thermal behaviour of H - H60.

| Composition | $T_{Cryst}$ (°C) | | $\alpha_{Cryst}$ (%) | | $T_{Melt}$ in °C | | CTE × $10^{-6}$/°C | CTE % reduction w.r.t H |
|---|---|---|---|---|---|---|---|---|
| | F | Pnt | F | Pnt | F | Pnt | | |
| H | 105.70 | 110.82 | 59.54 | 61.74 | 131.47 | 130.88 | 135±3.29 | --- |
| H20 | 112.67 | 113.12 | 49.12 | 50.72 | 132.51 | 131.24 | 106±3.85 | 21.48 |
| H40 | 112.92 | 113.23 | 33.71 | 37.01 | 130.45 | 131.29 | 88±2.65 | 34.81 |
| H60 | 112.59 | 113.27 | 25.79 | 28.59 | 130.86 | 130.90 | 75±1.15 | 44.44 |

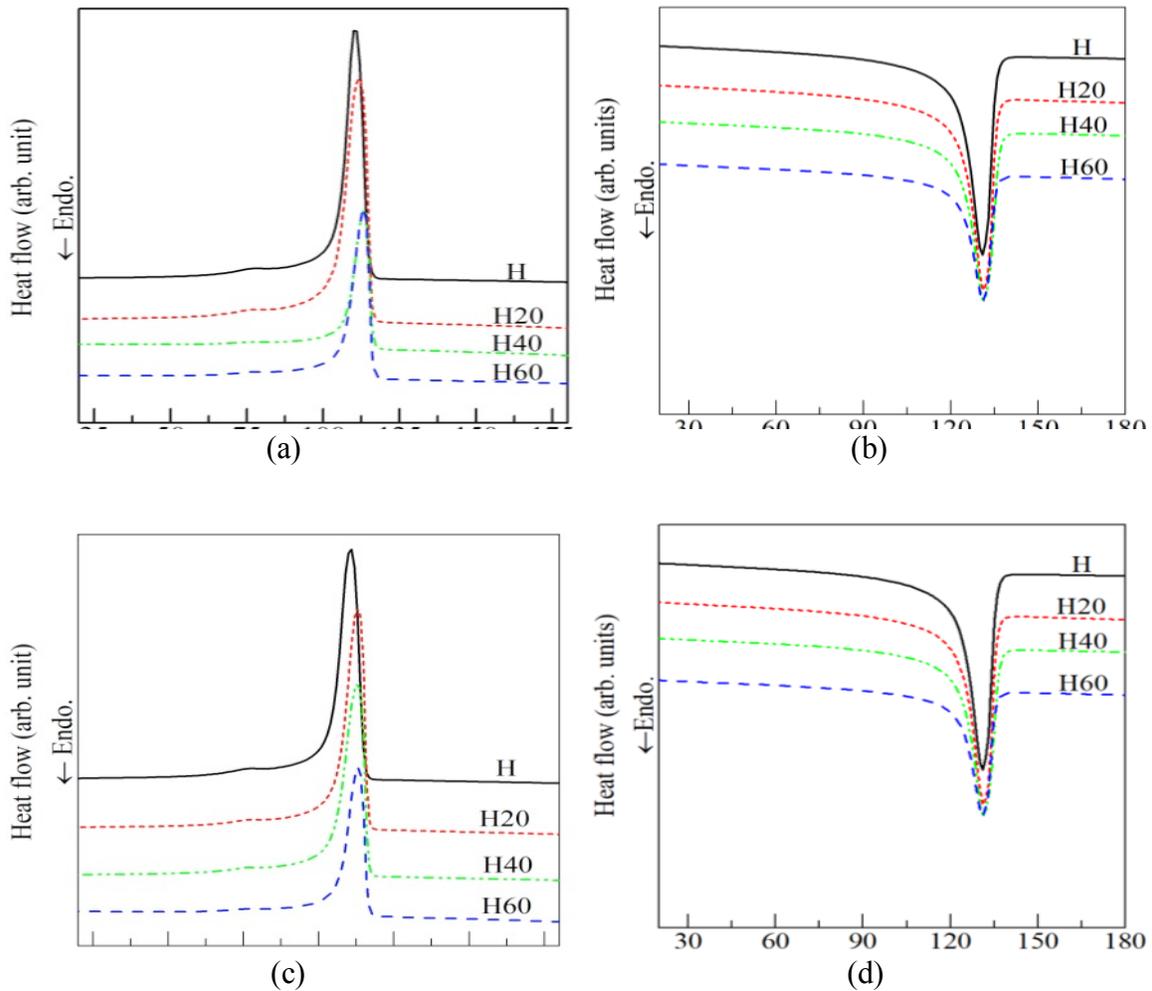

Figure 5. DSC for crystallization peaks: Cooling cycle in (a) filaments and (c) prints. Melting peaks from heating cycle (2$^{nd}$) in (b) filaments and (d) prints.

Printed samples also show similar behavior where $\alpha_{Cryst}$ dropped from 61.74 (H) to 28.59% (H60). Compared to filaments, the corresponding prints have a higher $\alpha_{Cryst}$ and is anticipated to provide higher dimensional stability and reduce warpage related issues. Extruded filaments are subjected to sort of quenching as it passes through the water bath immediately after the extrusion. Thereby, a very little time and energy is available for melt crystallization of filaments [64, 65] as compared to prints wherein samples cool slowly within the printer chamber. Due to the resistance offered by GMB to the flow of polymer chain $\alpha_{Cryst}$ decreases in foams along with the reduction of the crystal domain of H [32, 66, 67]. Hence dimensionally stable foam prints without any warpage can be successfully 3D printed having a potential weight saving of ~28% (Table 2).

*Investigations on filaments tensile properties*

The tensile response is governed by the dispersion of reinforcement, filler size, matrix interaction, and inherent properties of the matrix [68]. In order to use filament as feedstock material in the 3D printer, it must meet specific requirements like shape retention without excessive bending to absorb frictional forces while going through drive rollers [69]. Bending can be avoided by keeping the filament rigid enough to withstand the push of drive roller without damaging the associated printer elements. Figure 6 presents tensile stress-strain plots of filaments.

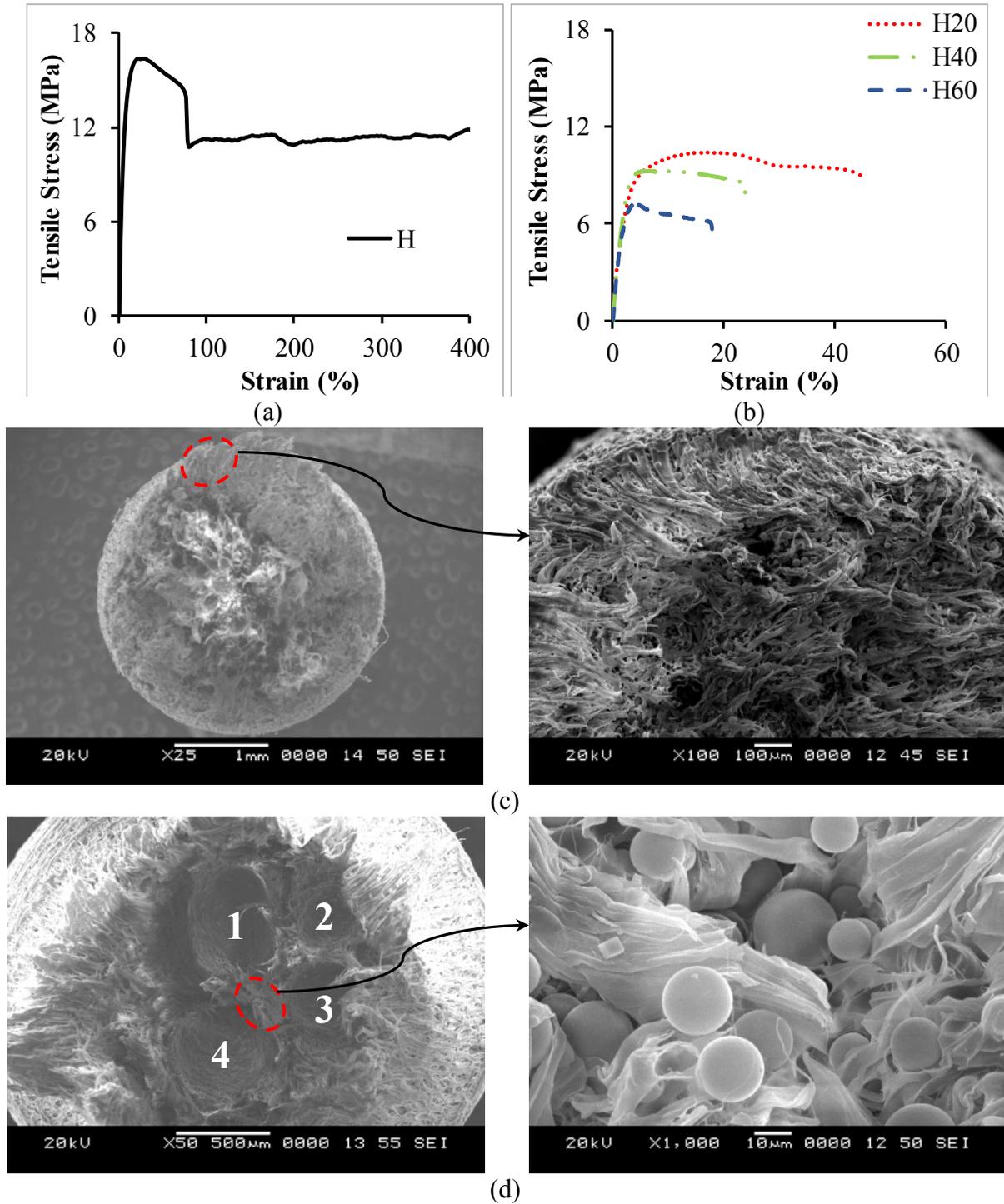

Figure 6. Representative filament stress-strain plot of (a) H and (b) H20 - H60. SEM of (c) H20 and (d) H60 filament post tensile tests.

Stiffer intact GMB particles increase filament modulus by 8.17, 14.40, and 46.81% in H20, H40, and H60, respectively, as compared to H (Table 6 and Figure 6b). HDPE filament is strained to more than 1000 % without any breakage due to its ductility. However, only up to 400% strain is graphed in Figure 6a. H40 and H60 failed within ~25% strain, as seen from Figure 6b. H20 exhibits more than 40% strain with the highest UTS of 12.63 MPa among foams. A higher amount of matrix in H20 resists the tensile load effectively by the plastic deformation of the entire cross-section, as observed from Figure 6c. The marked area in Figure 6c shows the formation of a new surface at bulk scale, enhancing strain.

H60 has the highest void content of 7.7% (Table 2) among foams resulting in much earlier filament fracture due to a reduction in the effective area arising from elongated pores coalescence (locations 1 - 4 in Figure 6d). Nevertheless, H60 exhibits the highest modulus because of a higher number of intact GMB particles (marked area in Figure 6d). Strength decreases with increasing filler content because of weaker bonding between GMB and HDPE, as seen in Figure 4c. Further, with increasing GMB content, HDPE volume decreases, lowering the ductile phase substantially, resulting in lower strength values. Filament strength can be increased by surface treating of GMB particles that lead to enhanced interfacial bonding, which is not within the scope of this work. Such a surface treatment approach needs careful attention as coupling agents increase brittleness and can hamper spooling flexibility. The focus of the current work is the development of lightweight composite foam filaments for 3DP using as received constituent materials, so that the processing time and cost is minimum, and enahnces industrial adaptability for components where modulus and comparable strength are the design criteria.

*3D printing of GMB/HDPE*
All the samples are printed in rectilinear pattern having print orientation in Y-axis. A layer thickness of 0.32 mm is set to provide adequate clearance between the nozzle and the printed part. A printing speed of 35 mm/s is kept constant for all the samples to improve the surface finish and lower the warpage. Infill is kept at 100% to achieve structural stability in addition to comparative analysis with fully dense injection and compression molded samples. A multiplier is set to 1 and 1.2 for H - H40 and H60, respectively, based on the MFI estimations. Up to 60% MFI reduction, layers are deposited without any difficulties with 35 mm/s printing speed. With reducing MFI above 60%, blocking of the nozzle is experienced, and hence a multiplier of 1.2 was set for H60 for a given nozzle temperature setting. Nozzle temperature is set above the Vicat softening point (124 °C) of HDPE. The printing and bed temperatures are varied across two temperature settings of 225, 245 °C and 80, 100 °C, respectively for the reasons mentioned in the earlier section.

The experiments are carried out based on 3DP of HDPE as it exhibits maximum warpage compared to foams. Nonetheless, with the suitable printing parameters of H, H20 - H60 samples are also printed. Table 4 and Figure 7 show the observations pertaining to the experimental tests carried out to identify suitable printing parameters. All H - H60 compositions are printed on the HDPE plate for optimum bonding between the first print layer with the base. Table 4 and Figure 7 shows that the printing and bed temperatures of 245 and 80 °C respectively are best suited for quality printing of H - H60. Hence, all the samples are printed using the parameters as listed in Table 5.

Table 4. Experimental test of 3D printing parameters.

| Printing temperature (°C) | Print bed temperature (°C) | Observation | Figure 7 |
|---|---|---|---|
| 225 | 80 | Improper layer deposition, Interlayer defects | a b |
| 225 | 100 | Merging of the bottom layer with plate | c |
| 245 | 80 | Proper layer deposition, Absence of interlayer defects, Easier removal of print from the plate, No warpage | d |
| 245 | 100 | Maximum warpage, Defective part | e |

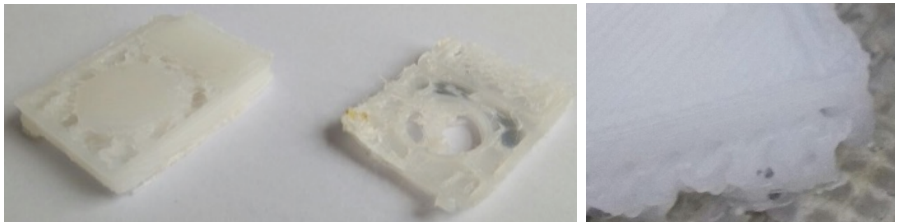

(a)          (b)

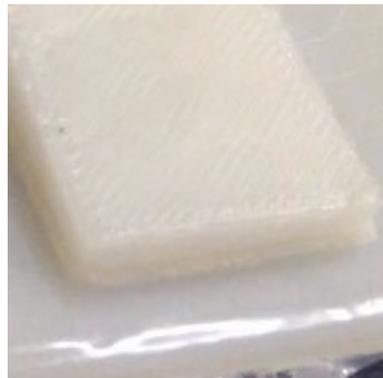 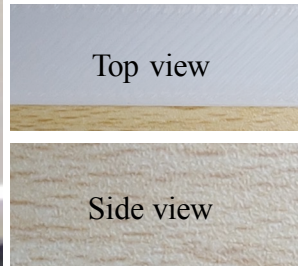

(c)          (d)

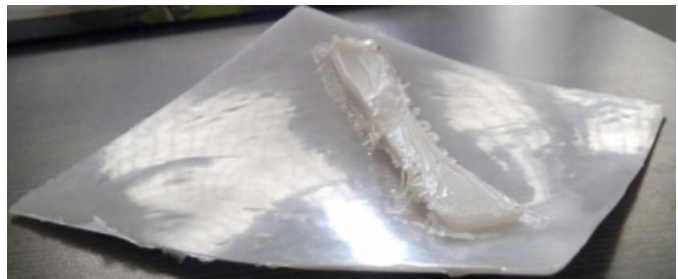

(e)

Figure 7. Challenges in 3D printing of HDPE. The description of various parts is presented in Table 4.

Table 5. Printing values utilized in current work.

| Parameters | Value |
|---|---|
| Temp. of Nozzle (°C) | 245 |
| Printing bed temp. (°C) | 80 |
| Layer height (mm) | 0.32 |
| Extrusion Multiplier | H - H40: 1, H60: 1.2 |
| Print speed (mm/sec) | 35 |
| Print pattern | Rectilinear |
| Orientation of part | Y-axis |
| Infill (%) | 100 (±45° to x-axis) |

3D printed samples are cooled within the build chamber till room temperature is reached. Printed samples exhibit consistent bonding between the layers with the least warpage (Figure 8a). The marked area in Figure 8a indicates very good seamless diffusion between the layers at higher magnification. This fact reaffirms the suitability of printing parameters utilized in the present work (Table 5). Micrographs of freeze-fractured HDPE print show very few voids (Figure 8b), while H60 micrograph (Figure 8c) exhibits uniform GMB dispersion along with elongated voids. Such elongated voids at higher filler volume % are the result of lower MFI and reduced melt viscosity. Printed samples have more void content compared to filaments (Table 2) because of air gaps between adjacent raster (Figure 8d). Air gaps are observed to be increasing with GMB content due to lower matrix phase, higher melt viscosity, and reduced CTE values. Such air gaps might enhance damping and compressive capabilities as mentioned earlier.

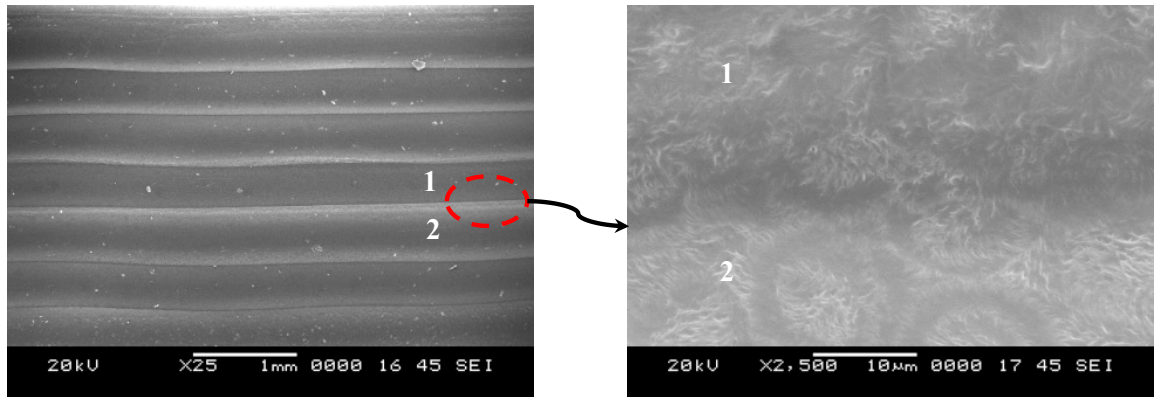

(a)

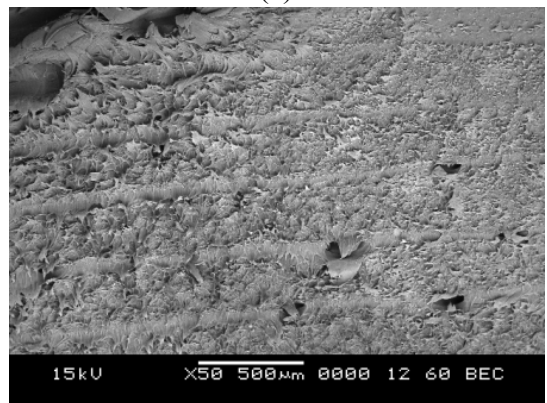

(b)

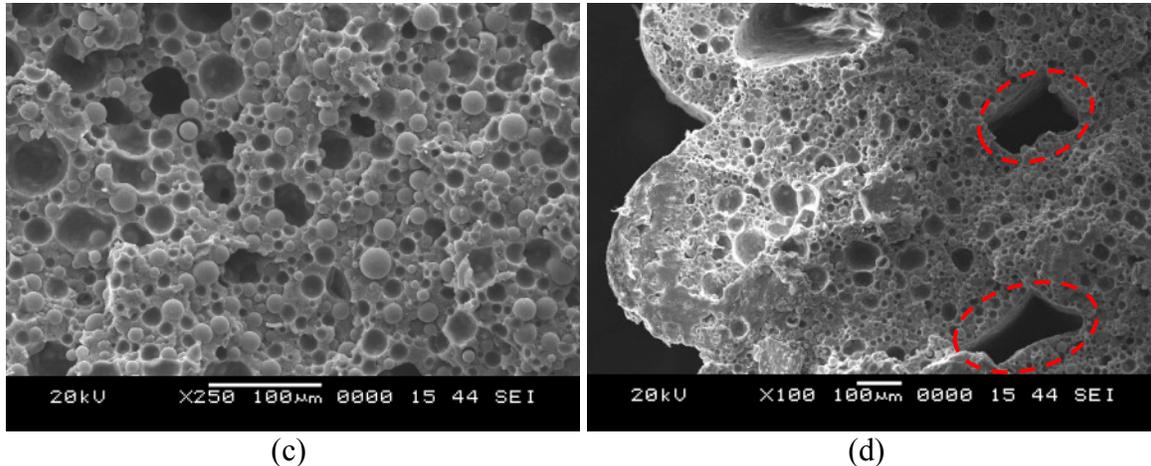

(c)                              (d)

Figure 8. Micrograph of printed (a) H in thickness direction and (b) freeze fractured across the thickness (c) H60 and (d) associated raster gaps in H60.

*CTE of prints*
The addition of GMB in the HDPE matrix lowers CTE, as seen from Table 3 [70, 71]. At higher printing temperatures, dimensional stability can be achieved by adding GMB into HDPE. This indicates that the warpage can be avoided to a greater extent in printed components with dimensional stability and lower residual thermal stresses [70]. The entrapped gas inside the hollow GMB offers resistance against heat flow, which results in lower thermal conductivity. Also, there is a large difference in the CTE values of both the constituents, which is GMB and HDPE. Further, CTE also helps in understanding the raster diffusion mechanism and air gap formation in 3DP. Warpage, which is clearly evident from Figure 7e is a crucial and challenging factor while printing neat HDPE due to higher CTE values. Nonetheless, appropriate printing and bed temperatures can effectively address this issue. Dimensionally stable prints are observed in foams due to lower CTE as a result of the presence of lower thermal conductive gases within hollow GMB that limits the flow of heat [72, 73]. Among foams, H60 print showed the lowest CTE leading to minimal raster diffusion, thereby resulting in air gaps (Figure 8d). Such air gaps make syntactic foams lighter (~2 – 4%), as seen from Table 2.

*Tensile response of prints*
A similar trend of stress-strain response is exhibited by 3D printed H - H60 samples as in filaments, and the values are listed in Table 6. Breakage of pure HDPE filament is not seen even after a strain of up to 1000%, while HDPE print could sustain only up to ~45% strain, indicating a behavioral change from ductile to brittle phase post 3DP. HDPE is extruded twice, once during filament formation, and secondly in printer nozzle extruder. Such multiple extrusion cycles result in polymer chain alignment, associated crosslinking due to thermal processing leading to the hardening process. Failure strain for 3D printed H40, and H60 foams are 21.66 and 14.49%, respectively, whereas H20 shows up to 30.48% strain. In the case of HDPE, a long necking region is clearly observed (Figure 9a) due to raster fibrillation resulting in broom-like fibrous ends. Such fibrous ends are a result of new surface formations because of extensive plastic deformation (micrograph of the marked area in Figure 9a). H40 and H60 foam prints show no necking region and fracture in a typical brittle manner, which is also seen from the fractographic area wherein matrix plastic deformation is hardly seen (Figure 9b).

Table 6. Filament and prints tensile response.

| Composition | Modulus in MPa | | Ultimate Tensile Strength (MPa) | | Elongation at Ultimate Tensile Strength (%) | | Fracture Strength (MPa) | | Fracture strain in % | |
|---|---|---|---|---|---|---|---|---|---|---|
| | F | Pnt | F | Pnt | F | Pnt | F | Pnt | F | Pnt |
| H | 722 ±16.73 | 810.25 ±16.73 | 16.4 ±0.22 | 17.68 ±0.21 | 17.90 ±0.26 | 15.04 ±0.23 | --- | 6.68 ±0.11 | --- | 93.00 ±1.03 |
| H20 | 781 ±17.95 | 865.56 ±17.79 | 10.45 ±0.42 | 12.8 ±0.35 | 12.63 ±0.33 | 5.68 ±0.29 | 8.93 ±0.23 | 10.39 ±0.29 | 44.27 ±0.23 | 30.48 ±0.10 |
| H40 | 826 ±14.27 | 1125.68 ±12.41 | 9.25 ±0.39 | 9.49 ±0.49 | 5.27 ±0.35 | 3.11 ±0.31 | 7.01 ±0.19 | 8.24 ±0.25 | 23.81 ±0.22 | 21.66 ±0.06 |
| H60 | 1060 ±18.53 | 1199.26 ±11.53 | 7.16 ±0.17 | 8.45 ±0.18 | 2.39 ±0.21 | 4.69 ±0.11 | 5.90 ±0.14 | 7.78 ±0.19 | 16.53 ±0.31 | 14.49 ±0.07 |

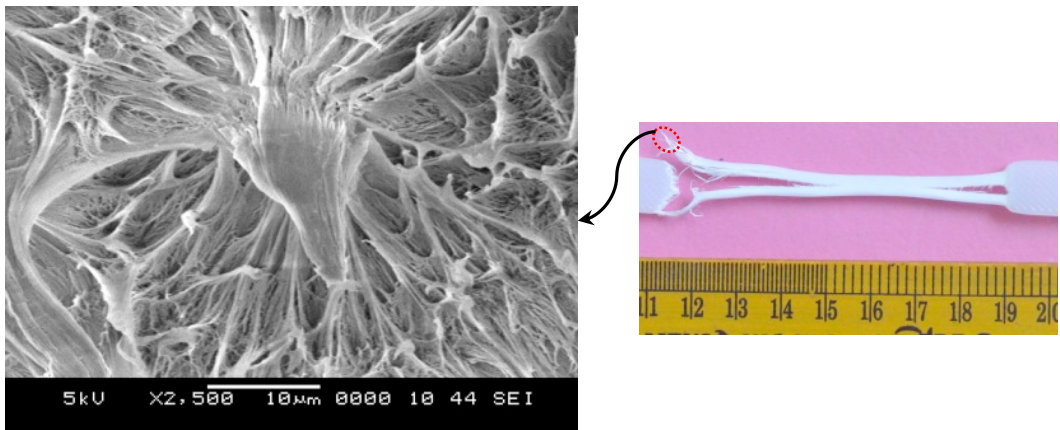

(a)

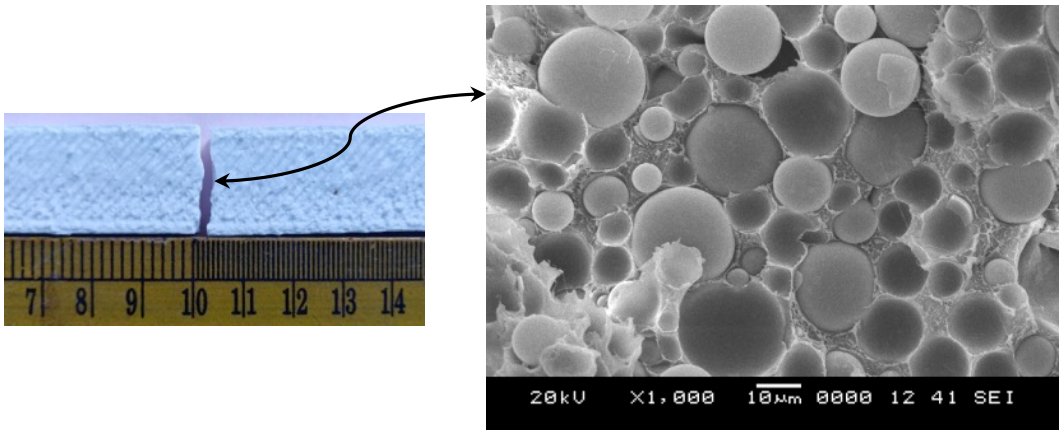

(b)

Figure 9. Fractographic analysis of representative 3D printed (a) H and (b) H60 post tensile test.

All the microballoons are observed to be intact, signifying potential weight saving of ~28% (Table 2) is successfully achieved post printing. Intact GMB particles at higher filler % make matrix responsible for load carrying, which succumbs early owing to induced brittleness post printing. Comparative analysis between the filament and printed coupons pertaining to modulus and strength show an increase by 12.22, 10.83, 36.28, 13.14%, and 7.8, 22.49, 2.59, 18.02%, respectively. GMB/HDPE prints results are compared with injection molded cenosphere/HDPE foams. 3D printed HDPE shows appreciable UTS with a higher elastic modulus of 53.17% when compared with injection molded foams. 3D printed foam specimen elongation at UTS and fracture strength are 47.45% and ~3 times higher than that of injection molded specimen [37]. Modulus of foam increases with GMB % (Table 6). Among foams, H60 displays the highest modulus and is 48.02% higher than HDPE print. 3D printed H - H60 registered 1.5 - 1.8 times higher modulus than molded counterparts with zero tooling cost. Foam prints fracture strength is 1.16 - 1.56 times higher when compared with H. For weight-sensitive applications, specific properties of foams are essential since printing allows flexibility in developing integrated (joint less) components with complex designs. Among all foams, H60 and H20 exhibit the highest specific modulus and strength, respectively. Table 7 shows the GMB/HDPE weight saving potential through estimations of $E/\rho^n$ (n = 1, 2 and 3). Values in Table 7 clearly indicated that 3D printed GMB/HDPE foams can be used effectively in buoyancy modules, automotive and aerospace components of integrated complex designs.

Table 7. Weight saving quantification parameters of H and their foams.

| Composition | $\frac{E}{\rho}$ MPa/kg/m³ | $\frac{E}{\rho^2}$ MPa/(kg/m³)²×10⁻³ | $\frac{E}{\rho^3}$ MPa/(kg/m³)³×10⁻⁶ |
|---|---|---|---|
| H | 0.87 | 0.94 | 1.02 |
| H20 | 1.05 | 1.27 | 1.54 |
| H40 | 1.51 | 2.02 | 2.71 |
| H60 | 1.80 | 2.69 | 4.02 |

*Flexural behavior of prints*
Foams displayed brittle fracture (Figure 10a) as compared to HDPE, which did not fail until 10% strain (Figure 10b). GMB inclusion induces brittleness in the compliant HDPE matrix. Crack initiated from the tensile side and propagated along the loading direction until it meets the compressive side. This is a typical flexural failure mode. It is interesting to note here that, the crack did not propagate along the deposited layers, confirming again the suitable printing parameters (Table 5) chosen for printing. As seen in Table 8, intact GMB particles (Figure 10c) increases the modulus with higher filler loadings. H60 modulus is 1.37 times higher than the H, while strength is observed to be decreased due to poor interface bonding between constituent elements and rasters gaps (Figure 11).

Table 8. Flexural response of H – H60 prints.

| Composition | Modulus in MPa | Strength in MPa | Fracture strength in MPa | Fracture strain in % | Specific modulus in MPa/kg/m³ | Specific strength in MPa/kg/m³×10⁻³ |
|---|---|---|---|---|---|---|
| H | 990±11.28 | 25.4±0.12 | --- | --- | 1.068 | 27.40 |
| H20 | 1210±19.56 | 21.0±0.58 | 20.34±0.32 | 6.88±0.09 | 1.465 | 25.42 |
| H40 | 1280±11.87 | 17.1±0.47 | 16.89±0.41 | 6.04±0.11 | 1.716 | 22.92 |
| H60 | 1360±11.23 | 15.1±0.72 | 15.00±0.79 | 3.15±0.07 | 2.036 | 22.60 |

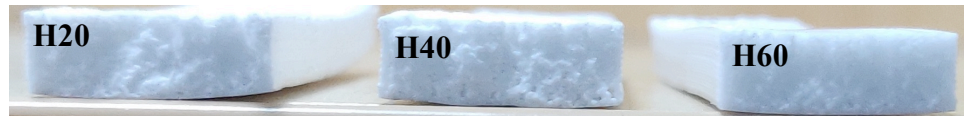

(a)

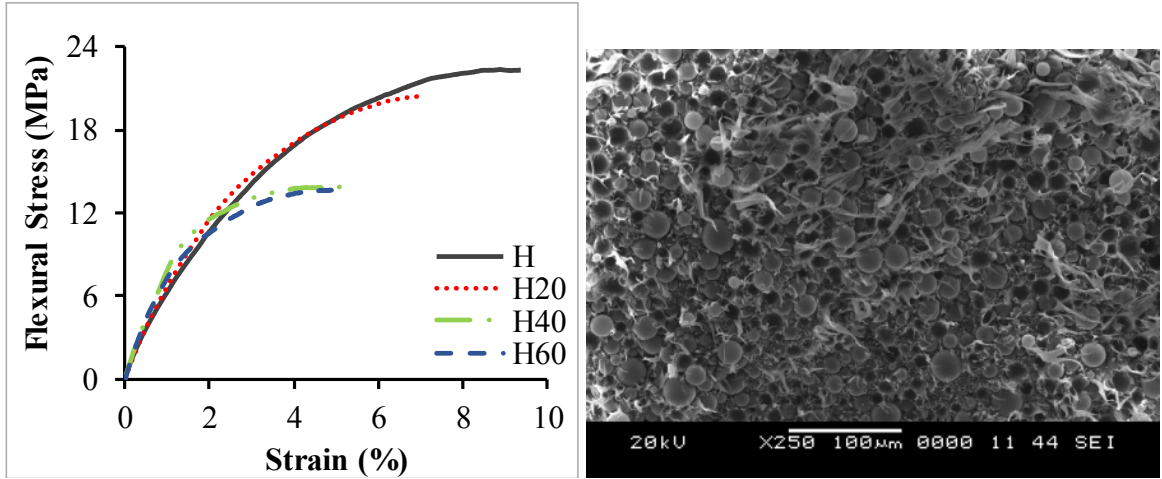

(b)                                    (c)

Figure 10. (a) Fractured foam samples post flexural test. Representative (b) stress-strain plots for prints and (c) H60 micrograph post flexure test.

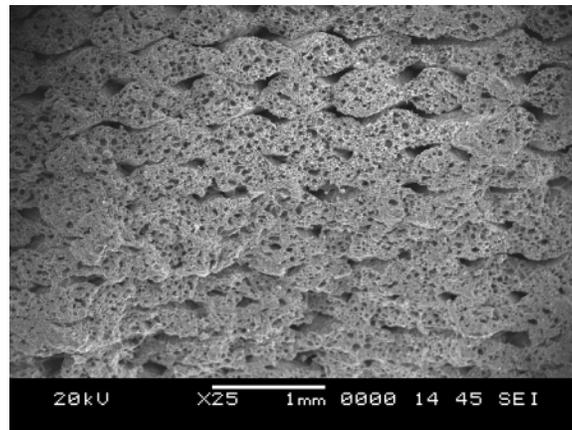

Figure 11. 3D printed representative H60 micrograph showing raster gaps.

GMBs embedded in the HDPE matrix increases the specific modulus by ~2 times compared to H. Modulus of H - H60 printed foams is higher by 1.39 - 1.08 times against molded counterparts, whereas strength is observed higher and comparable in case of H and H20 foams respectively. Drop-in strength by 1.14 and 1.27 is noted for printed H40 and H60 respectively against fully dense molded samples and is obvious owing to higher matrix porosity resulting from raster gaps [74]. With increasing filler loadings, these raster gaps volume increases due to lower CTE values. Nevertheless, these gaps can be minimized by overlapped deposition of layers and will be explored in future investigations. Tensile and flexural strength is observed to be decreasing as constituent materials are used in as received condition, as mentioned earlier. Furthermore, filler addition increases amorphous fraction leading to more restrained matrix flow and polymer chain mobility resulting in weaker interfaces. Enhancing the bonding between the constituents through appropriate coupling agents might increase the strength but at the expense of a substantial reduction in ductility, which may hamper filament extrusion and the 3DP process.

**Property Graphs**

Figure 12 and Figure 13 [2, 26, 37, 75] show the tensile and flexural response respectively as a function of composite density fabricated using different processing routes. Hollow particle filled composites exhibit promising properties to be exploited in weight sensitive applications as compared to solid filled material systems. The density of GMB based 3D printed foams is in between injection and compression molded foams. Tensile modulus of GMB based 3D printed composites outperform injection and compression molded composite (Figure 12a) except wood filled composites. The strength of GMB based printed foams is comparable to compression and injection molded samples (Figure 12b). Flexural modulus of GMB based 3D printed composites is greater than other syntactic foams realized by conventional manufacturing processes (Figure 13a). Flexural strength is comparable to composites produced from compression and injection molding (Figure 13b). Figure 14 helps materials designers/scientists and industrial practitioners in deciding the composition of 3D printed GMB/HDPE foam based on the properties required as per the envisaged application. The choice of suitable extrusion and printing conditions with no particle breakage results in substantial weight reduction by ~28%, as shown in the current work. Such a weight reduction for complex integrated 3D printed components would enhance the performance with reduced carbon footprints.

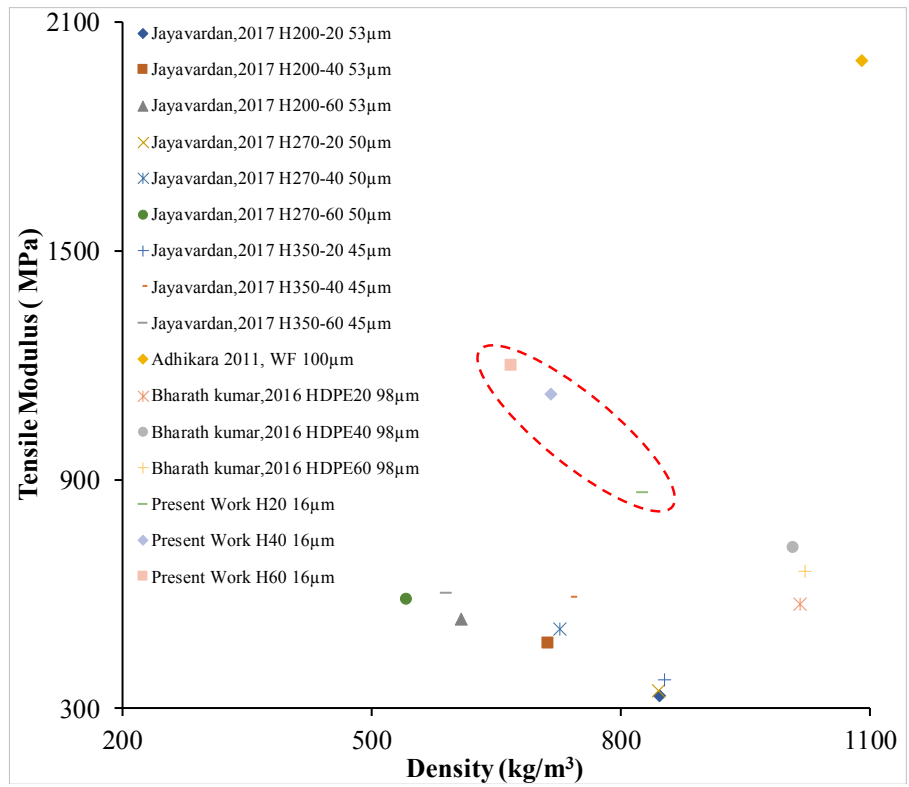

(a)

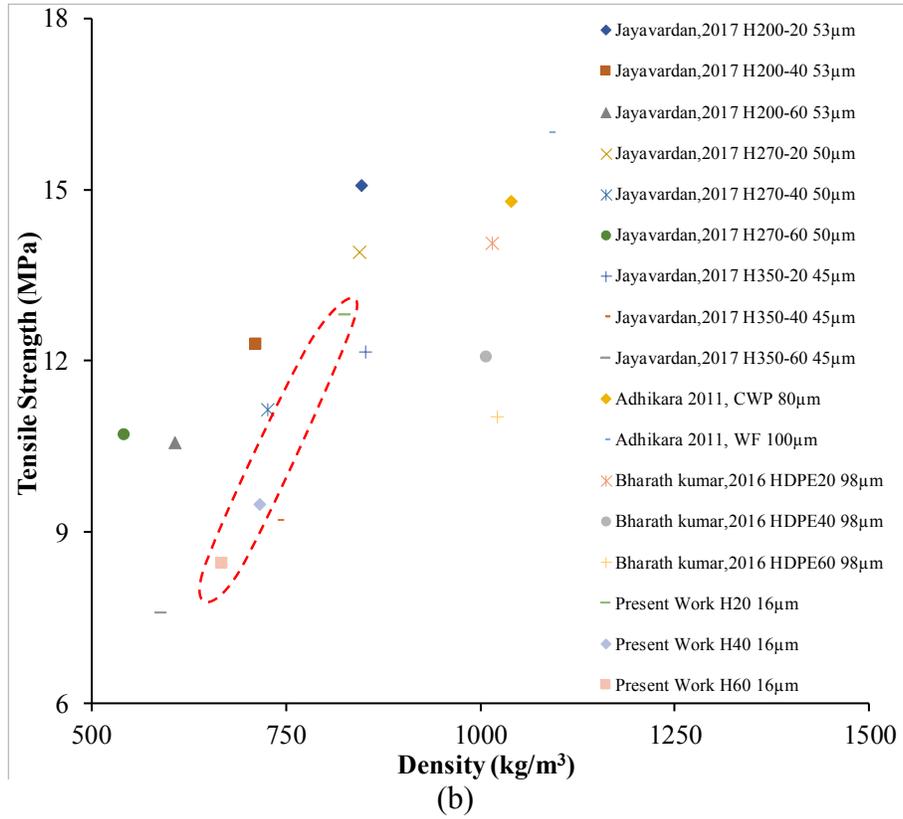

(b)

Figure 12. Tensile (a) modulus and (b) strength of HDPE composite [2, 26, 37, 75].

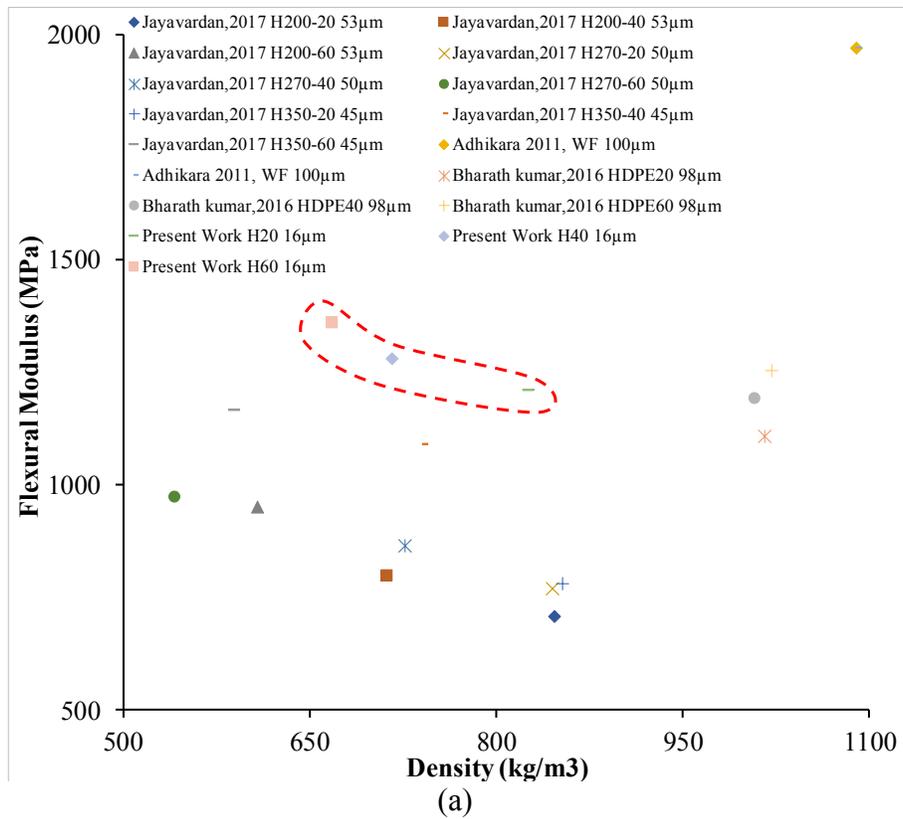

(a)

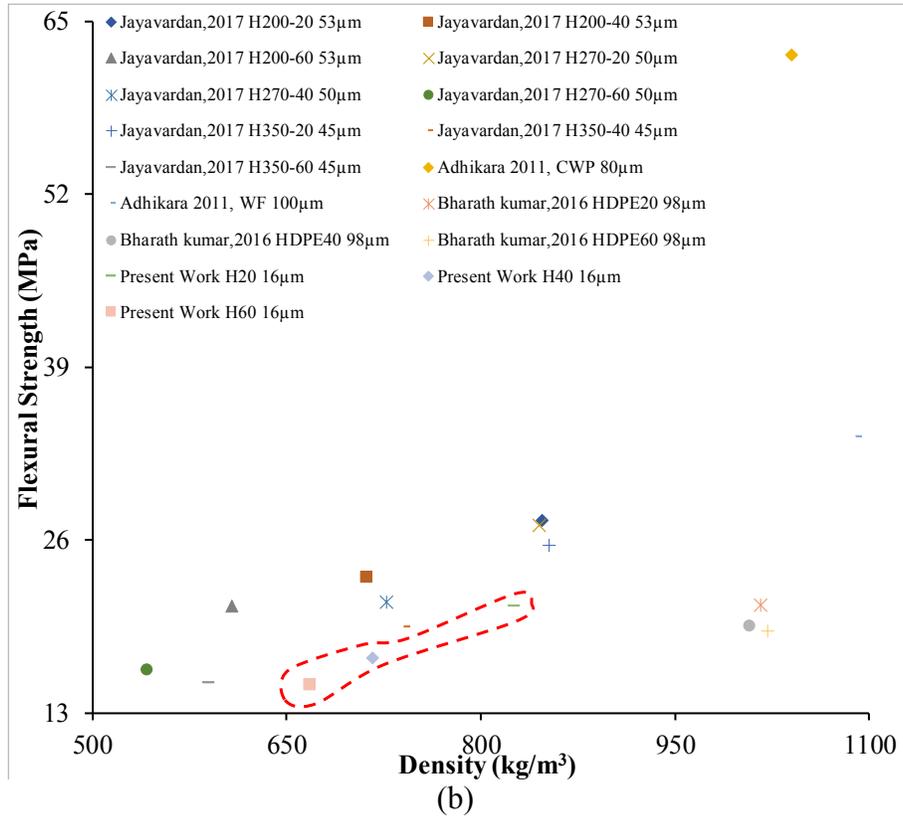

(b)

Figure 13. Flexural (a) modulus and (b) strength of HDPE composite [2, 26, 37, 75].

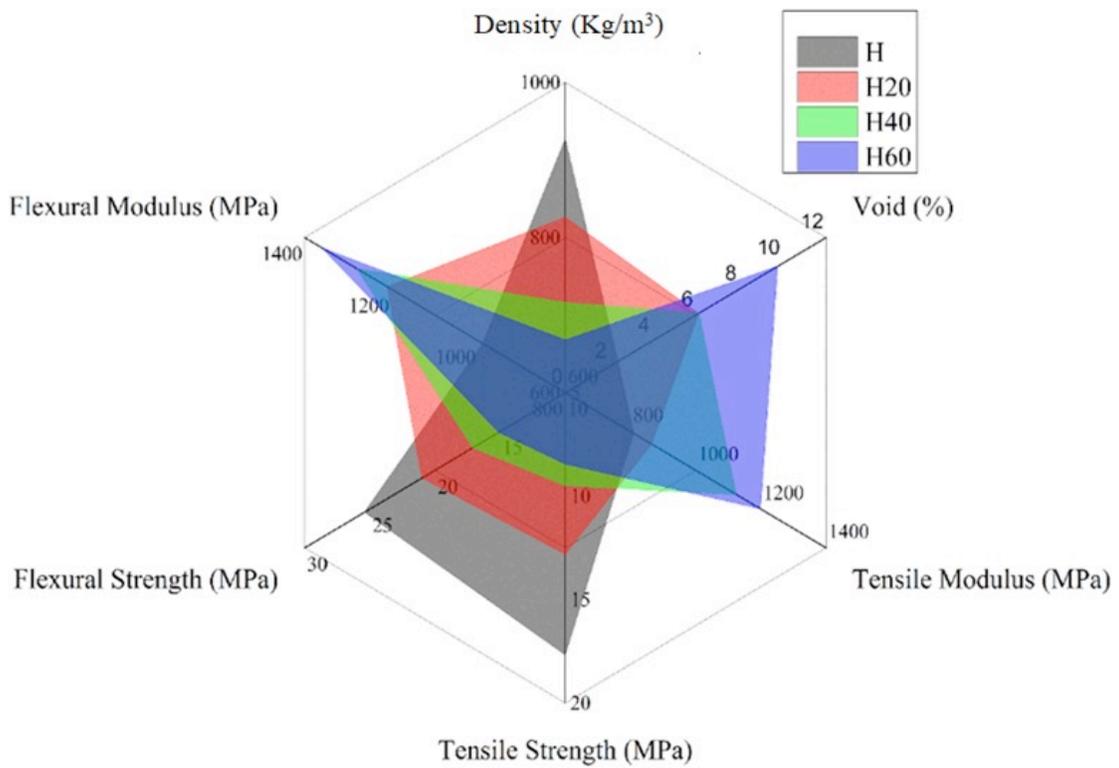

Figure 14. Comparative chart of the 3D printed GMB/HDPE properties.

**Conclusion**
GMB based lightweight composite foam feedstock is successfully synthesized to be used on a commercial printer for weight-sensitive applications. Filaments and 3D printed samples are tested for mechanical characterization to check their adaptability and feasibility for three-dimensional printing applications, and a summary of the results is presented below:
- Void contents increase in filaments and prints by 0.84-7.70% and 2.42-9.73%, respectively, with increasing filler content. 3D printed foams exhibit a 3 phase foam structure.
- An increase in GMB content decreases the MFI of HDPE.
- Loss modulus, storage modulus, and complex viscosity increase with increasing GMB content. At lower frequency, complex viscosity is maximum and decreases as the frequency increases showing shear thinning behavior in GMB/HDPE blends. Loss and storage modulus showed an increasing trend with an increase in GMB % and frequency.
- Filler content has no significant effect on peak melting temperature ($T_{Melt}$) of filaments and prints. The degree of crystallinity ($\alpha_{Cryst}$) decreases by 56.68 and 53.69 % for foam filaments and prints, respectively, with increasing GMB content as compared to HDPE. Compared to filaments, the corresponding prints have a higher $\alpha_{Cryst}$ and is a positive sign towards dimensional stability and warpage related issues.
- Stiffer intact GMB particles increase the filament modulus by 8.17 - 46.81% in H20 - H60, respectively, as compared to H.
- The addition of GMB in HDPE decreases the CTE of prints substantially, making the prints dimensionally more stable.
- Among foams, H60 displays the highest modulus, which is 48.02% higher than HDPE print. 3D printed H - H60 registered 1.5 - 1.8 times higher modulus than molded counterparts. Printed H20 - H60 has 1.16 - 1.56 times higher fracture strength than the printed H.
- GMBs embedded in the HDPE matrix increases the specific modulus by ~2 times compared to H. Modulus of H - H60 printed foams is higher by 1.39 - 1.08 times against molded counterparts.
- 3D printed GMB/HDPE foams having substantial weight saving potential (28%) with superior specific mechanical properties and reduced carbon footprints are successfully realized.

Current work successfully demonstrated the development of lightweight feedstock filament with an intention to widen available material choices for commercially available 3D printers. GMB/HDPE integrated complex geometrical components can be printed without any warpage, as presented in this work. Strength enhancement can be realized by surface modification of the constituent materials along with the strategy of overlapping raster and is the focus of future investigations.

**Acknowledgment**
The authors wish to acknowledge the support by SPARC grant (SPARC/2018-2019/P439/SL), Govt. of India. Mechanical Engineering Department of National Institute of Technology, Karnataka, Surathkal, is thanked for providing the facilities and support. The authors would also like to acknowledge the support by U.S. Office of Naval Research - Young Investigator Program (ONR-YIP) award [Grant No.: N00014-19-1-2206] for conducting the research presented here.